\documentclass[lettersize,journal,comsoc]{IEEEtran}
\input{befe}

\begin{document}
	
	\title{Diffusion-based Generative Prior for Low-Complexity MIMO Channel Estimation}
	\author{Benedikt Fesl,~\IEEEmembership{Graduate Student Member,~IEEE,} Michael Baur~\IEEEmembership{Graduate Student Member,~IEEE,} Florian Strasser, Michael Joham,~\IEEEmembership{Member,~IEEE,} and Wolfgang Utschick,~\IEEEmembership{Fellow,~IEEE}
		\thanks{
			The authors are with Lehrstuhl f\"ur Methoden der Signalverarbeitung, Technische Universit\"at M\"unchen, 80333 M\"unchen, Germany  (e-mail: benedikt.fesl@tum.de; mi.baur@tum.de; florian1.strasser@tum.de; joham@tum.de; utschick@tum.de).
		}
	}
	
	
	
	\maketitle

	\begin{abstract}
        This work proposes a novel channel estimator based on \acp{dm}, one of the currently top-rated generative models. Contrary to related works utilizing generative priors, a lightweight \ac{cnn} with positional embedding of the \ac{snr} information is designed by learning the channel distribution in the sparse angular domain. Combined with an estimation strategy that avoids stochastic resampling and truncates reverse diffusion steps that account for lower \ac{snr} than the given pilot observation, the resulting \ac{dm} estimator has both low complexity and memory overhead. Numerical results exhibit better performance than state-of-the-art channel estimators utilizing generative priors.
	\end{abstract}
	

\begin{figure}[b]
\onecolumn
\centering
\copyright \scriptsize{This work has been submitted to the IEEE for possible publication. Copyright may be transferred without notice, after which this version may no longer be accessible.}
\vspace{-1.3cm}
\twocolumn
\end{figure}
\section{Introduction}
\IEEEPARstart{G}{enerative} models have shown great success in learning complex data distributions and subsequently leveraging this prior information for wireless communication applications. 
This success is built on the great importance of inferring knowledge of the unknown and generally complex channel distribution of, e.g., a \ac{bs} environment through a representative dataset.
Consequently, the development of advanced channel estimation methodologies has ensued, primarily relying on state-of-the-art generative models such as \acp{gmm} \cite{9842343}, \acp{mfa} \cite{fesl2023lowrank}, \acp{gan} \cite{9252921}, or \acp{vae} \cite{baur2023leveraging}.

Recently, \acp{dm} \cite{Ho20_DDPM} and score-based models \cite{song2020generative} have been identified among the most powerful generative models. Both models are closely related by learning the data distribution through corrupting clean samples with additive (Gaussian) noise and learning the reverse process to generate new samples from pure noise.
However, the huge computational overhead associated with these models, i.e., a large number of \ac{nn} forward passes together with resampling after each step in the reverse process, makes their direct application difficult in a real-time application like channel estimation.

Nevertheless, \acp{dm} have been used in wireless communications, e.g., for channel coding \cite{channel_coding} and joint source-channel coding \cite{wu2023cddm}.
The work in \cite{MIMO_channel_estimation_diff} proposed to utilize a score-based model to perform channel estimation through posterior sampling. However, the approach has several disadvantages that hinder its usage in practical applications, e.g., a high number of network parameters in the order of millions, a huge number of reverse steps in the order of thousands that have to be evaluated online, and the stochasticity of the estimate.

Recently, the work in \cite{fesl2024icml} proposed a deterministic denoising strategy that utilizes a \ac{dm} where the observation's \ac{snr} level is matched with the corresponding timestep of the \ac{dm}, drastically reducing the necessary number of reverse steps without requiring resampling. Moreover, this denoising strategy was shown to be asymptotically \ac{mse}-optimal if the number of total diffusion timesteps grows large. For practical distributions, it was shown that already a moderate number of timesteps is sufficient for achieving a strong denoising performance close to the utopian bound of the \ac{cme}.
Despite these advantages, the deployed \ac{nn} based on a sophisticated architecture with parameters in the order of millions might still be impractical to be used for channel estimation. Furthermore, \ac{mimo} channels contain structural properties, e.g., sparsity in the angular/beamspace domain, that can be leveraged to design lightweight \acp{nn} of reduced sizes. 

In this work, we provide the following contributions:
\begin{itemize}
\item We propose a novel channel estimator with the currently top-rated \ac{dm} as generative prior.
\item To account for the shortcomings in terms of computational complexity, memory overhead, and robustness in the reverse process, we design a \ac{dm}-based channel estimator with a lightweight \ac{cnn} by learning the channel distribution in the sparse angular domain. 
\item We highlight the connection of the resulting \ac{dm}-based channel estimator to the \ac{mse}-optimal \ac{cme} and its asymptotic optimality. 
\item We evaluate the proposed \ac{dm}-based estimator on different channel models, showing its versatile applicability. Furthermore, even though the presented \ac{dm} has a drastically reduced number of parameters and online computational complexity as compared to existing techniques based on generative priors, it exhibits a better estimation performance.
\end{itemize}

\section{Preliminaries}
\subsection{MIMO System Model}
Consider an uplink transmission where a \ac{mt} equipped with $\Ntx$ transmit antennas sends $N_{\text{p}}$ pilots to the \ac{bs} with $\Nrx$ receive antennas, yielding the received signal
\begin{align}
    \B Y = \B H \B P + \B N \in \mathbb{C}^{\Nrx \times N_{\text{p}}}
    \label{eq:syst_mod}
\end{align}
where $\B H\in \mathbb{C}^{\Nrx\times \Ntx}$ is the wireless channel matrix following an unknown distribution, $\B P \in \mathbb{C}^{\Ntx \times N_{\text{p}}}$ is the pilot matrix known to the receiver, and $\B N$ is \ac{awgn} with $N_{\text{p}}$ columns $\B n_i\sim \mathcal{N}_{\mathbb{C}}(\B 0, \eta^2\eye)$. In the remainder of this work, we consider $N_{\text{p}} = \Ntx$, i.e., full pilot observations and a unitary pilot matrix, e.g., a \ac{dft} matrix, fulfilling $\B P\B P\h = \eye$. 
By lowercase bold symbols, we refer to the vectorized expressions of \eqref{eq:syst_mod}, e.g., $\B h =\vect(\B H)$, which we use interchangeably.
We consider $\mathbb{E} [\B h] = \B 0$ and $\mathbb{E}[\|\B h\|_2^2] = \Nrx\Ntx$, which can be ensured by pre-processing, allowing to define the \ac{snr} as $1/ \eta^2$.

\subsection{Diffusion Model as Generative Prior}
We briefly review the formulations of \acp{dm} from \cite{Ho20_DDPM} with vectorized expressions for notational convenience.
Given a data distribution $\B h = \B h_0 \sim p(\B h_0)$, the \ac{dm}'s \textit{forward process} produces \textit{latent variables} $\B h_1$ through $\B h_T$ by adding Gaussian noise at time $t$ with the hyperparameters $\alpha_t \in (0,1)$ such that
\begin{align}
    \B h_{t} &= \sqrt{\alpha_{t}} \B h_{t-1} + \sqrt{1-\alpha_t} \B \varepsilon_{t-1} 
    \label{eq:xt_repara}
    \\
    &= \sqrt{\balpha_{t}} \B h_{0} + \sqrt{1-\balpha_t} \B \varepsilon_0
    \label{eq:xt_repara_x0}
\end{align}
with $\B \varepsilon_i \sim \NC(\B 0, \eye)$ and $\balpha_t = \prod_{i=1}^t \alpha_i$.  
The \textit{reverse process} is, exactly like the forward process, a Markov chain but with parameterized Gaussian transitions
\begin{align}
    p_\btheta(\B h_{t-1} | \B h_t) = \NC(\B h_{t-1}; \B \mu_{\btheta}(\B h_t, t), \sigma^2_t\eye),
    \label{eq:reverse_transition}
\end{align}
motivated by the fact that the forward and reverse processes of a Markov chain have the same functional form \cite{sohl_dickstein15}.
The transitions in \eqref{eq:reverse_transition} are generally intractable to compute; thus, they are learned via a \ac{nn} through the variational inference principle by utilizing the forward posteriors, which are tractable when conditioned on $\B h_0$ \cite{Ho20_DDPM}, i.e.,
\begin{align}
    q(\B h_{t-1} | \B h_{t}, \B h_0) &= \NC(\B h_{t-1}; \tilde{\B \mu}(\B h_t, \B h_0), \sigma_t^2\eye),
    \label{eq:ground_truth_x0}
\\
    \tilde{\B \mu}(\B h_t, \B h_0) &= \frac{\sqrt{\bar{\alpha}_{t-1}} (1 - \alpha_t)}{1-\bar{\alpha}_t} \B h_0 + \frac{\sqrt{\alpha_t} (1 - \bar{\alpha}_{t-1})}{1-\bar{\alpha}_t} \B h_t,
    \label{eq:cond_mean_ground_truth}
    \\
    \sigma_t^2 &= \frac{(1-\alpha_t)(1 - \bar{\alpha}_{t-1})}{1 - \bar{\alpha}_t}.
    \label{eq:cond_var_ground_truth}
\end{align}
Since the conditional variance $\sigma_t^2$ is a time-dependent constant, a \ac{nn} is trained to parameterize the conditional mean $\B \mu_{\btheta}(\B h_t,t)$ from \eqref{eq:reverse_transition} for a given latent $\B h_t$ at its input. We denote this \ac{nn} function as $f_{\btheta,t}^{(T)}(\B h_t) := \B \mu_{\btheta}(\B h_t,t)$.
The training of the \ac{dm} is performed by maximizing the \ac{elbo} on the log-likelihood $\log p(\B h_0)$, cf. \cite{Ho20_DDPM} for a detailed derivation.
The \ac{dm} timesteps can be equivalently interpreted as different \ac{snr} steps by defining the \ac{dm}'s \ac{snr} of step $t$ as
\begin{align}
    \text{SNR}_{\text{DPM}}(t) = \frac{\mathbb{E}[\|\sqrt{\bar{\alpha}_t} \B h_0\|_2^2]}{\mathbb{E}[\|\sqrt{1 - \bar{\alpha}_t} \B \varepsilon_0\|_2^2]} 
    = \frac{\bar{\alpha}_t}{1 - \bar{\alpha}_t},
    \label{eq:dm_snr}
\end{align}
cf. \eqref{eq:xt_repara_x0}, which monotonically decreases for increasing $t$.

\section{Channel Estimation}\label{sec:channel_est}
The optimal channel estimator in terms of the \ac{mse} is the \ac{cme} \cite[Appendix A.3]{tse_fundamentals}, defined as
\begin{align}
    \mathbb{E}[\B h | \B y]
    = \int \B h p(\B h | \B y) \op d\B h
   =  \int \B h \frac{p(\B y | \B h) p(\B h)}{p(\B y)}\op d\B h.
\end{align}
The \ac{cme} is generally highly nonlinear and intractable to compute if the prior distribution $p(\B h)$ is unknown. Thus, we utilize the \ac{dm} as generative prior to parameterize a channel estimator that has implicitly learned the prior distribution.

\subsection{Diffusion-based Channel Estimator}\label{subsec:estimator}
Wireless communication channels have unique structural properties that differ from \textit{natural signals} in other domains, e.g., images. 
A well-known property is that the channel can be transformed into the angular/beamspace domain representation via a Fourier transform \cite[Sec. 7.3.3]{tse_fundamentals}. 
In massive \ac{mimo}, the angular domain representation is sparse or highly compressible, especially if the number of multipath propagation clusters and the angular spread is low, which is the case, e.g., in mmWave communications.
As a consequence, learning the channel distribution in the sparse angular domain is likely to allow the \ac{dm} to need fewer parameters, which, in turn, makes the training and inference faster and more stable. However, we note that we do not make any specific assumption about the sparsity level or the structure of the channel distribution.

Therefore, we transform the channels from a given training dataset $\mathcal{H} = \{\B H_m\}_{m=1}^{M_{\text{train}}}$ with $M_{\text{train}}$ training samples into the angular domain as $\tilde{\mathcal{H}} = \{\tilde{\B H}_m = \fft(\B H_m)\}_{m=1}^{M_{\text{train}}}$, where $\tilde{\B H}_m = \fft(\B H_m)$ represents the channels in the angular domain, which can be implemented utilizing two-dimensional \acp{fft}. 
Afterward, the \ac{dm} is trained offline in the usual way by maximizing the \ac{elbo}, cf. \cite{Ho20_DDPM}.

\begin{algorithm}[tb]
\caption{Channel estimation via a DM as generative prior.}
\label{alg:reverse_process}
\begin{algorithmic}[1]
\renewcommand{\algorithmicensure}{\textbf{Offline DM Training Phase}}
\ENSURE
\REQUIRE Training dataset $\mathcal{H} = \{\B H_m\}_{m=1}^{M_{\text{train}}}$
\STATE Transform into angular domain $\tilde{\mathcal{H}} = \{\fft(\B H_m)\}_{m=1}^{M_{\text{train}}}$
\STATE Train DM $\{f_{\btheta,t}^{(T)}\}_{t=1}^T$ with $\tilde{\mathcal{H}}$, cf. \cite{Ho20_DDPM}
\par\vskip.5\baselineskip\hrule height .4pt\par\vskip.5\baselineskip
\renewcommand{\algorithmicensure}{\textbf{Online DM-based Channel Estimation Phase}}
\ENSURE
\REQUIRE $\{f_{\btheta,t}^{(T)}\}_{t=1}^T$, $\B Y$, $\eta^2$, $\B P$
\STATE Compute LS estimate $\hat{\B H} \leftarrow \B Y \B P\h$
\STATE Normalize observation's variance $\hat{\B H} \leftarrow (1 + \eta^2)^{-\frac{1}{2}} \hat{\B H}$
\STATE Transform into angular domain $\hat{\B H} \leftarrow \fft(\hat{\B H})$
\STATE $\tinit = \argmin_t |\text{SNR}(\B Y) - \text{SNR}_{\text{DM}}(t)|$
\STATE Initialize \ac{dm} reverse process $\hat{\B H}_{\tinit} \leftarrow \hat{\B H}$
\FOR{$t=\tinit$ {\bfseries down to} $1$}
    \STATE $\hat{\B H}_{t-1} \leftarrow f_{\btheta,t}^{(T)}(\hat{\B H}_{t}) $
\ENDFOR
\STATE Transform back into spatial domain $\hat{\B H} \leftarrow \ifft(\hat{\B H}_{0} )$
\end{algorithmic}
\end{algorithm}

For the online channel estimation, we adopt the deterministic reverse process from \cite{fesl2024icml} with several changes. 
First, the pilot matrix is decorrelated by computing the \ac{ls} solution, i.e., 
\begin{align}
    \hat{\B H}_{\text{LS}} = \B Y \B P\h = \B H + \tilde{\B N}
    \label{eq:ls}
\end{align}
where $\tilde{\B N} = \B N \B P\h$ is \ac{awgn} with variance $\eta^2$ due to $\B P$ being unitary.
Assuming knowledge of the observation's \ac{snr}, the \ac{ls} estimate is normalized as $\hat{\B H}_{\text{init}} = (1 + \eta^2)^{-\frac{1}{2}} \hat{\B H}_{\text{LS}}$ since the deployed \ac{dm} is variance-preserving, cf. \eqref{eq:xt_repara}. Afterward, the observation is transformed into the angular domain via $\hat{\B H}_{\text{ang}} = \fft(\hat{\B H}_{\text{init}})$. Note that the noise distribution is unaltered by the unitary Fourier transformation. Next, the \ac{dm}'s timestep~$\tinit$ that best matches the observation's \ac{snr} is found by utilizing the \ac{snr} representation of the \ac{dm}, cf. \eqref{eq:dm_snr}, as
\begin{align}
    \tinit = \argmin_t |\text{SNR}(\B Y) - \text{SNR}_{\text{DM}}(t)|.
\end{align}
Subsequently, the \ac{dm}'s reverse process is initialized by setting $\hat{\B H}_{\tinit} = \hat{\B H}_{\text{ang}}$. As a consequence, the higher the observation's \ac{snr} is, the less \ac{dm} reverse steps have to be performed and, in turn, the lower the latency of the channel estimation.
This is in sharp contrast to the work in \cite{MIMO_channel_estimation_diff} where the inference process is initialized with i.i.d. Gaussian noise and a full reverse sampling process is employed, irrespective of the \ac{snr}. 

After initializing an intermediate \ac{dm} timestep, the stepwise conditional mean of \eqref{eq:reverse_transition} is iteratively forwarded, without drawing a stochastic sample from $p_{\btheta}(\B h_{t-1}|\B h_t)$, from $t=\tinit$ down to $t=1$, ultimately yielding an estimate of $\B H_0$. This can be denoted by the concatenation of the \ac{nn} functions 
\begin{align}
    \hat{\B H}_0 = f_{\btheta,1}^{(T)}( f^{(T)}_{\btheta, 2}( \cdots f^{(T)}_{\btheta, \tinit}(\hat{\B H}_\tinit) \cdots ))
    =  f^{(T)}_{\btheta, 1:\tinit}(\hat{\B H}_\tinit).
\end{align}
Finally, the resulting estimate is transformed back into the spatial domain via the two-dimensional inverse \ac{fft}, yielding $\hat{\B H} = \ifft(\hat{\B H_0})$. The complete offline training and online estimation procedures are concisely summarized in Algorithm~\ref{alg:reverse_process}.

\subsection{Asymptotic Optimality}\label{subsec:asymptotic_opt}
The work in \cite{fesl2024icml} has proved the convergence of the \ac{dm}-based estimator to the ground-truth \ac{cme} if the number of diffusion steps $T$ of the \ac{dm} grows large under the assumptions of a well-trained \ac{dm} and a specific asymptotic behavior of the \ac{dm}'s hyperparameters, cf. \cite[Th. 4.6]{fesl2024icml}. Since the pilot decorrelation and the Fourier transforms are invertible, we can conclude that the asymptotic behavior is also valid in the considered case, i.e., under the assumptions of \cite[Th. 4.6]{fesl2024icml} and for every given observation $\B Y$ it holds that
\begin{align}
    \limT \left\|\mathbb{E}[\B H |\B Y] - \ifft\left(f_{\btheta,1:\tinit}^{(T)}\left(\fft\left(\tfrac{1}{\xi}\B Y\B P\h\right)\right)\right) \right\| = 0
\end{align}
where $\xi = \sqrt{1 + \eta^2}$.

We note that the asymptotic behavior holds for any given \ac{snr} value of the observation; however, in a practical system, it can be reasonably assumed that the \ac{bs} receives pilots from a limited \ac{snr} range, raising the question of how many diffusion timesteps are practically necessary for a strong estimation performance. As we show in \Cref{sec:num_results}, already a moderate to small number of diffusion steps $T$, in contrast to the score-based channel estimator from \cite{MIMO_channel_estimation_diff}, is sufficient to outperform state-of-the-art generative prior-aided channel estimators in a massive \ac{mimo} system.

\subsection{Diffusion Model Network Architecture}\label{subsec:network_architecture}
Instead of utilizing a sophisticated \ac{nn} architecture commonly used for \acp{dm}, cf. \cite{Ho20_DDPM}, we design a lightweight \ac{cnn}. Enabled by the sparse structure of typical wireless channel distributions in the angular domain, cf. \Cref{subsec:estimator}, the lightweight design addresses the importance of low memory overhead and computational feasibility in real-time wireless channel estimation. 
The network architecture is detailed in Fig. \ref{fig:network_architecture}, which is explained in the following.
We use parameter sharing across all \ac{dm} timesteps, i.e., only a single network is deployed for all \ac{dm} timesteps. To this end, a Transformer sinusoidal position embedding of the time/\ac{snr} information is utilized to yield $\B t\in \mathbb{R}^{C_{\text{init}}}$, cf. \cite[Sec. 3.5]{vaswani2023attention} for details, which is, after going through a linear layer, subsequently split into a scaling vector $\B t_{\text{s}}\in \mathbb{R}^{C_{\text{max}}}$ and a bias vector $\B t_{\text{b}}\in \mathbb{R}^{C_{\text{max}}}$.

After stacking the real- and imaginary parts of the input $\hat{\B H}_t$ into two convolutional channels, we employ two 2D convolution layers with kernel size of $k=3$ in both dimensions and \ac{relu} activation, gradually increasing the number of convolution channels up to $C_{\text{max}}$, after which the time embedding is connected. Afterward, three 2D convolution layers map to the estimate $\hat{\B H}_{t-1}$ by gradually decreasing the convolutional channels. 
We choose the same linear schedule of $\alpha_t$ between constants $\alpha_1$ and $\alpha_T$ as in \cite[Table 1]{fesl2024icml}.
Further details and hyperparameters can be found in the publicly available simulation code.\footnote{\url{https://github.com/benediktfesl/Diffusion_channel_est}}

 \begin{figure}[t]
	\centering
    \includegraphics[width=\columnwidth]{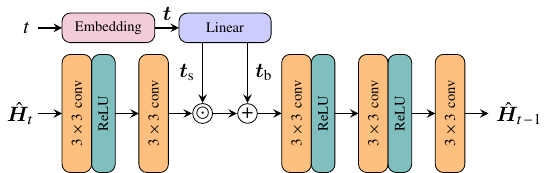}
    \vspace{-0.1cm}
	\caption{\ac{dm} architecture with a lightweight \ac{cnn} and positional embedding of the \ac{snr} information. The \ac{nn} parameters are shared across all timesteps.
	}
	\label{fig:network_architecture}
\end{figure}

\section{Numerical Results}\label{sec:num_results}
We consider a massive \ac{mimo} scenario with $(\Nrx, \Ntx) = (64, 16)$. For the \ac{dm}, we choose $C_{\text{init}}=16$ and $C_{\text{max}} = 64$.
All data-aided approaches are trained on $M_{\text{train}} = 100{,}000$ training samples and the \ac{mse} is normalized by $\Nrx\Ntx$.

\subsection{Channel Models}\label{subsec:channel_models}
We test all estimators on two different datasets from different channel models to allow a fair and unbiased evaluation.
First, we work with the \ac{3gpp} spatial channel model~\cite{3gpp,learning_mmse} where channels are modeled conditionally Gaussian.
The random vector $\B \delta$ collects the angles of arrival/departure and path gains of the main propagation clusters between a \ac{mt} and the \ac{bs} to construct the channel covariance matrix $\B C_{\B \delta}$, cf.~\cite[eq. (5)]{learning_mmse}. 
For every channel sample, we draw a new $\B \delta$ and subsequently draw the sample as $ \B h \sim \NC(\B 0, \B C_{\B \delta}) $, which results in an overall non-Gaussian channel distribution~\cite{9842343}.

Second, version 2.4 of the QuaDRiGa channel simulator~\cite{QuaDRiGa1} is used. 
We simulate an urban macrocell scenario at a frequency of 6 GHz.
The \ac{bs}'s height is 25 meters, covering a $120^\circ$ sector.
The distances between the \acp{mt} and the \ac{bs} are 35--500 meters, and we consider a \ac{los} scenario.
The \ac{bs} and the \acp{mt} are equipped with a uniform linear array with half-wavelength spacing.
The channels are post-processed to remove the effective path gain.

\subsection{Baseline Approaches}\label{subsec:baselines}
We compare the \ac{dm}-based channel estimator with the following classical and generative prior-aided baselines. The simple \ac{ls} solution in \eqref{eq:ls} is denoted by ``LS.'' A linear \ac{mmse} estimate based on the global sample covariance matrix $\B C = \frac{1}{M_{\text{train}}}\sum_{m=1}^{M_{\text{train}}} \B h_m\B h_m\h$ utilizing the training samples is computed as $\hat{\B h}_{\text{Scov}} = \B C(\B C + \eta^2\eye)\inv \B y$, labeled ``Scov''.
Assuming genie knowledge of the ground-truth covariance matrix $\B C_{\B \delta}$ from the 3GPP model allows to evaluate the estimator $\hat{\B h}_{\text{genie}} = \B C_{\B \delta}(\B C_{\B \delta} + \eta^2 \eye)\inv \B y$, which yields a lower bound for all estimators, cf. \cite{9842343}, labeled ``genie.''

We additionally compare the proposed \ac{dm} to the \ac{gmm}-based estimator from \cite{9842343} with either full covariance matrices with $K=128$ components (``GMM'') or the Kronecker version thereof with $(K_{\text{rx}}, K_{\text{tx}}) = (16, 8)$ components (``GMM Kron''), also yielding a total of $K=128$ components, cf. \cite{fesl_structured}.
We also evaluate the score-based channel estimator from \cite{MIMO_channel_estimation_diff} 
based on a convolutional RefineNet architecture with $D=6$ residual blocks with $L=7$ layers with $W=32$ channels after the first layer and a maximum channel size of $C_{\text{max}} = 128$, and a kernel size of $k=3$ in each dimension. 
After training, we perform a hyperparameter search on the test data and take the best \ac{mse} value of all reverse steps, yielding a genie-aided bound on its performance. For the present system and channel data, we have found $\alpha_0 = 10^{-8}$ and $\beta = 10^{-2}$, cf.~\cite[eq. (16)]{MIMO_channel_estimation_diff}, to yield the best performance. The corresponding curves are labeled as ``Score.'' Since the \ac{gan}-based estimator~\cite{9252921} is outperformed by the ``Score'' model \cite{MIMO_channel_estimation_diff}, it is not evaluated in this work due to space limitations.

\begin{table}[t]
\caption{Memory and complexity analysis for $(\Nrx, \Ntx) = (64, 16)$.}
\label{tab:memory}
\small
\begin{center}
\vspace{-0.3cm}
\renewcommand{\arraystretch}{1.0}
\begin{tabular}{||l|c|c||}
\hline
Method & Parameters & Online Complexity \\
\Xhline{4\arrayrulewidth}
Scov  & $1.05\cdot 10^6$ &  $\mathcal{O}(N^2_{\text{rx}}N^2_{\text{tx}})$\\ \hline
GMM  & $1.35 \cdot 10^8$  & $\mathcal{O}(KN^2_{\text{rx}}N^2_{\text{tx}})$  \\  \hline
GMM Kron  & $7.22\cdot 10^4$ & $\mathcal{O}(KN^2_{\text{rx}}N^2_{\text{tx}})$ \\ \hline 
Score  & $5.89\cdot 10^6$ & $\mathcal{O}(T_{\text{sc}}k^2DLC_{\text{max}}^2 \Nrx\Ntx)$ \\ \hline
DM (proposed) & $\mathbf{5.50} \boldsymbol{\cdot} \mathbf{10}^{\mathbf{4}}$ & $\mathcal{O}(\tinit k^2 C_{\text{max}}^2 \Nrx \Ntx)$  \\
\hline
\end{tabular}
\end{center}
\vskip -0.1in
\end{table}

\subsection{Complexity and Memory Analysis}\label{subsec:memory_complexity}
The number of necessary model parameters, determining the memory overhead, and the computational online complexity for the channel estimation are analyzed in Table \ref{tab:memory}. 
The ``Scov'' estimator requires $N^2_{\text{rx}}N^2_{\text{tx}}$ parameters, which scales badly in massive \ac{mimo} systems. Additionally, pre-computation of the individual filters for each \ac{snr} value is necessary to achieve the stated order of complexity, which might be intractable to achieve in practice due to memory issues. 
The ``GMM'' model, comprising $K$ covariance matrices of size $\Nrx\Ntx\times \Nrx\Ntx$, has an intractably high number of parameters, thus being prone to overfitting, as seen later. Therefore, the stated order of complexity of the $K$ linear \ac{mmse} filters with pre-computed inverses for each \ac{snr} value as proposed in~\cite{9842343} may not be realistic to achieve in massive \ac{mimo}. Although the ``GMM Kron'' version needs much fewer model parameters, it has the same online complexity, cf. \cite{fesl_structured}, and thus the same memory issues to enable parallelization and precomputation.
The ``Score'' model from \cite{MIMO_channel_estimation_diff} also has a high number of network parameters. 
In addition, the online complexity, dominated by a huge number of forward passes through a sophisticated deep \ac{nn}, is very high, detrimentally affecting the computational complexity and latency.

In comparison to the discussed baselines, the proposed \ac{dm} with $T=100$ total timesteps unifies a low memory overhead with several orders fewer parameters than the discussed baselines. In addition, it only has a linear scaling in the number of receive- and transmit-antennas. Due to the truncation of inference steps depending on the \ac{snr}, the number of forward passes $\tinit < T \ll T_{\text{sc}}$ is much lower than in the directly comparable score-based model \cite{MIMO_channel_estimation_diff}. An evaluation of $\tinit$ is further shown in Fig. \ref{fig:plot_t}.
Together with the great potential for parallelization on GPUs due to exceptionally low memory overhead, this highlights the excellent scaling properties of the proposed \ac{dm} in massive \ac{mimo} applications, achieving a practicably viable online complexity.

\begin{figure}[t]
	\centering
    \includegraphics{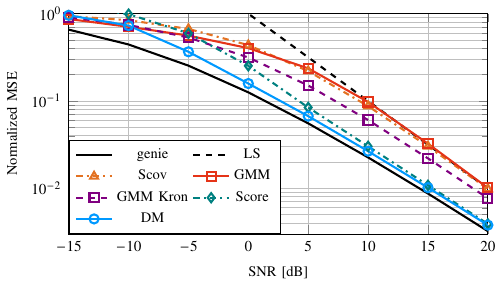}
	\caption{\ac{mse} performance for the \ac{3gpp} channel model with three propagation clusters and $T=100$ \ac{dm} timesteps.}
    \label{fig:mse_3gpp}
\end{figure}

\subsection{Performance Evaluation}\label{subsec:performance_eval}
Fig. \ref{fig:mse_3gpp} assesses the \ac{mse} performance for the \ac{3gpp} model, cf. \Cref{subsec:channel_models}, with three propagation clusters. 
The best estimator over the whole \ac{snr} range is the proposed \ac{dm} with $T=100$ timesteps, achieving an estimation performance close to the utopian bound ``genie'' and thus showing that the asymptotic optimality from \Cref{subsec:asymptotic_opt} is achievable already with a moderate number of timesteps $T$.
The ``Score'' model, although being comparably good as the \ac{dm} in the high \ac{snr} regime, suffers in performance in the low \ac{snr} regime. The ``GMM'' variant is highly overfitting due to the immense number of model parameters, cf. Table \ref{tab:memory}; in contrast, the ``GMM Kron'' version performs better, but with a considerable gap to the \ac{dm} estimator, especially in the high \ac{snr} regime.

In Fig. \ref{fig:mse_quadriga}, the \ac{mse} performance is evaluated for the QuaDRiGa model, cf. \Cref{subsec:channel_models}. The qualitative results are similar to those of Fig. \ref{fig:mse_3gpp}; however, the ``Score'' and the ``GMM Kron'' methods are performing almost equally well in this case.
Moreover, the \ac{dm} has an even larger gap to the baseline approaches with up to $5\operatorname{dB}$ gain in \ac{snr} compared to the ``Score'' model. In combination with the low memory overhead and computational complexity, cf. Table \ref{tab:memory}, this highlights the high potential of the proposed \ac{dm} estimator in practical applications.

Fig. \ref{fig:plot_T} and Fig. \ref{fig:plot_t} evaluate the \ac{mse} for the QuaDRiGa model for varying total \ac{dm} timesteps $T$ and over the intermediate \ac{dm} channel estimates in the reverse process from $t=\tinit$ to $t=0$, respectively. On the one hand, it can be observed that already a low number of \ac{dm} timesteps $T$ is sufficient for reasonable performance, saturating beyond $T=100$ for all \ac{snr} values. 
On the other hand, the intermediate channel estimates of the \ac{dm}'s reverse process are almost monotonically improving in performance, validating the stable reverse process. Furthermore, it can be observed that the \ac{dm} with $T=100$ timesteps converges to the same \ac{mse} as a \ac{dm} with $T=300$, demonstrating that a moderate number of timesteps $T$ is sufficient for a strong estimation performance.

\begin{figure}[t]
	\centering
    \includegraphics{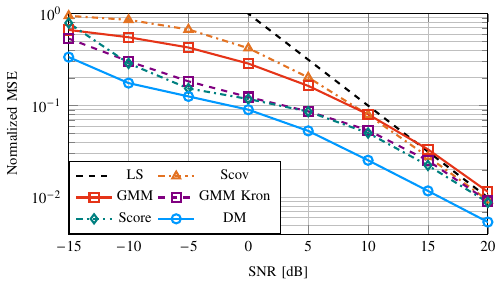}
	\caption{\ac{mse} performance for the QuaDRiGa channel model and $T=100$ \ac{dm} timesteps.}
     \label{fig:mse_quadriga}
\end{figure}

\begin{figure}[t]
	\centering
	\includegraphics{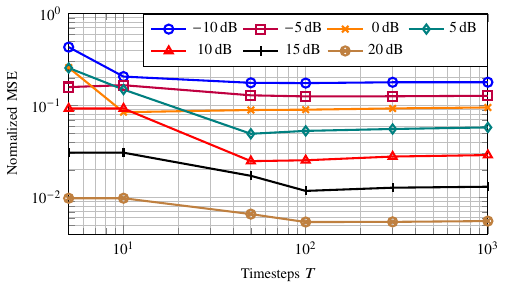}
	\caption{\ac{mse} performance over the total number of \ac{dm} timesteps $T$ for the QuaDRiGa model.}
     \label{fig:plot_T}
\end{figure}

\begin{figure}[t]
	\centering
	\includegraphics{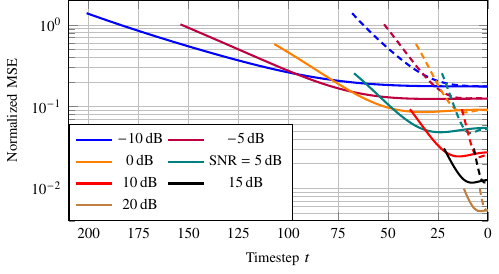}
	\caption{\ac{mse} performance of the intermediate estimates of the \ac{dm}'s timesteps for $T=300$ (solid) and $T=100$ (dashed) for the QuaDRiGa model.}
 \label{fig:plot_t}
\end{figure}

\section{Conclusion}
This letter introduced a novel \ac{mimo} channel estimator based on the \ac{dm} as generative prior. It has been shown that through learning the channel distribution in the highly compressible angular domain and employing an estimation strategy that has a lower latency toward higher \ac{snr} values, the proposed \ac{dm}-based estimator unifies low memory overhead together with low computational complexity, in addition to better estimation performance compared to state-of-the-art estimators based on generative priors.

\bibliographystyle{IEEEtran}
\bibliography{IEEEabrv,biblio}

\end{document}